\documentclass[preprint]{iopart}
\usepackage{graphicx}
\usepackage[a4paper,margin=1in]{geometry}
\usepackage{color}
\expandafter\let\csname equation*\endcsname\relax
\expandafter\let\csname endequation*\endcsname\relax
\usepackage{amsmath}
\usepackage{multirow}
\usepackage{epsfig}
\usepackage{epstopdf}
\usepackage{amssymb}
\usepackage{cancel}
\usepackage{float}
\usepackage{nameref}
\usepackage[utf8]{inputenc}
\usepackage{textcomp}
\usepackage{multicol}
\usepackage{indentfirst}
\usepackage{tabularx}
\usepackage{longtable}
\usepackage{setspace}
\usepackage{framed}
\usepackage{subcaption}
\usepackage{soul}
\usepackage{sidecap}
\usepackage{upgreek}
\usepackage{booktabs}
\usepackage{comment}
\usepackage{cuted}       
\usepackage{caption}     
\usepackage[sorting=none, backend=biber, style=numeric-comp, sortcites=true]{biblatex}
\newcommand{\myref}[1]{Figure~\ref{#1}}
\newcommand{\ignore}[1]{}

\usepackage[deletedmarkup=xout]{changes}

\usepackage[breaklinks=true,colorlinks=true,linkcolor=blue,urlcolor=blue,citecolor=blue]{hyperref}

\begin{document}


\title[]{Electronegativity effects on plasma dynamics in He/O$_2$ RF microplasma jets at atmospheric pressure}

\author{L. Vogelhuber$^{1}$, I. Korolov$^{2}$, M. Vass$^{2,3}$, K. Nösges$^{2}$, T. Bolles$^{2}$, K. Köhn$^{1}$, M. Klich$^{1}$, R. P. Brinkmann$^{1}$, T. Mussenbrock$^{2}$}

\address{$^1$Chair of Theoretical Electrical Engineering, Ruhr-University Bochum, D-44780, Bochum, Germany}
\address{$^2$Chair of Applied Electrodynamics and Plasma Technology, Ruhr-University Bochum, D-44780, Bochum, Germany}
\address{$^3$Institute for Solid State Physics and Optics, HUN-REN Wigner Research Centre for Physics, Budapest
1121, Hungary}

\date{\today}
\begin{abstract}
This work investigates the transitions between ohmic mode and Penning‐Gamma mode in a capacitively coupled radio frequency micro atmospheric pressure plasma jets (CCRF $\mu$APPJ) operated in He/O$_2$ mixtures by comparing phase‐resolved optical emission spectroscopy (PROES) measurements of helium excitation with numerical simulations. The simulations employ a hybrid model that treats electrons kinetically via PIC/MCC, while ions and neutrals are modeled fluid dynamically. These results reveal that increasing electronegativity causes inhomogeneities in the bulk electric field, consequently modulating electron impact excitation dynamics. A good agreement was found between experiments and simulations.
\end{abstract}

\section{Introduction}
Atmospheric pressure plasmas have attracted notable interest due to their wide range of applications, from medical and surface treatment \cite{Plasmas_for_medicine, PlasMedRev, MicroplasmaJetBioMEd, BioMedLaroussi} to CO$_{\rm{2}}$  conversion \cite{CO2Bogaerts, Co2ConvHana, Co2Conv}. Reactive nitrogen and oxygen species (RONS) play a crucial role in many technical and medical applications, such as surface treatment, bacterial inactivation, and wound healing \cite{WoundHealingPlasmaJet, SurfacePawlat_2016, SurfaceIchikiEtching, DepositionBabayan1998, Surface2article, ROSBacteria, WoundHealingRONS}. Specifically, cancer cells are more sensitive to plasma-induced RONS than normal cells, leading to selective cell death in tumors \cite{Hamouda2021, Kim2016}. Cold atmospheric pressure plasmas offer non-invasive treatment options by delivering RONS through plasma activated media, thus minimizing potential side effects from direct plasma exposure \cite{Jo2020}. Plasma jets effectively generate these reactive species, and therefore, the interest in micro atmospheric pressure plasma jets has increased, and various concepts have been proposed and studied \cite{PlasmaJetV1,kINPen, Barman_2020}. 

One example is the COST (Cooperation in Science and Technology) reference Jet (COST-Jet). It was developed as a standardized tool for investigating the effects of atmospheric pressure plasmas \cite{COSTRef_Golda_2016}. It has been extensively characterized through experiments \cite{CorrigendumCharacterizationCOST, Riedel_CostReproducibility, COSTRef_Golda_2016, GasandHeatDynamicsGolda,effluentKOROLOV} and numerical simulations \cite{Klich_2022,Klich_2025, Youfan0D, Liu_2023,Liu_2021}. The COST-Jet is often operated with helium and a small fraction of molecular admixtures such as oxygen or nitrogen \cite{COSTRef_Golda_2016,Liu_2021ModeTransition,Liu_2023,Liu_2021,Youfan0D}. 

Many experimental and numerical investigations have been conducted to understand the generation of RONS and the plasma dynamics inside such jets and in front of the effluent \cite{Waskoenig_2010AtomicOxigen, Korolov_2020HeMEta, Bischoff_2018TimeModulation, Hemke_2011Jet, Liu_2021ModeTransition}.
Numerical and experimental studies also demonstrate the crucial role of helium metastables in sustaining discharges \cite{RoleOFHEliumMEtaCostJet, Korolov_2020HeMEta, Bischoff_2018TimeModulation}. In molecular He/N$_2$ and He/O$_2$ mixtures, the metastable He energy level ($>\,19\,\mathrm{eV}$) exceeds the ionization thresholds of N$_2$ ($\sim15.6\,\mathrm{eV}$) and O$_2$ ($\sim12.1\,\mathrm{eV}$), allowing metastable helium to perform Penning ionization of molecular species. Due to the low fraction of molecular species, most electron impact processes occur with helium, making Penning ionization the primary ionization pathway at low N$_2$/O$_2$ concentrations \cite{MartensNitrogenDominant}. It was shown experimentally that in electronegative He/O$_2$ discharges, electron heating modes, namely, Penning-Gamma mode and ohmic mode, can transition from one to another depending on the admixture concentration and applied voltage \cite{Bischoff_2018TimeModulation,Liu_2021ModeTransition}. 

The underlying physical effects have been studied using numerical fluid dynamic simulations, which generally capture the trends but show deviations in the detailed spatiotemporal dynamics of helium excitation and the overall input power required to sustain the discharge compared to experimental observations \cite{Liu_2021ModeTransition, Hemke_2011Jet}. Strong local electric fields cause a high-energy tail in the electron energy distribution function (EEDF), deviating from a Maxwellian distribution \cite{Bischoff_2018TimeModulation, Liu_2021,Liu_2021ModeTransition}. Previous work has shown that accurately representing the EEDF in these jets and the resulting electron impact processes requires a kinetic description, as two‐term approximations in fluid dynamical simulations cannot reproduce all features observed experimentally \cite{ Liu_2021ModeTransition, Vass_2024_2DHybr}. The particle-in-cell/Monte Carlo collision (PIC/MCC) method allows for accurate resolution of electron dynamics in capacitively coupled radio frequency (CCRF) plasmas. This technique is widely applied in low-pressure plasmas \cite{Vass_2022PIC, PICBecker_2017, BastiTutorial} and atmospheric pressure plasmas \cite{Kawamura_2014, PICDonkó_2018, PICAtmo}. However, at atmospheric pressure, the high gas density and collision frequency numerically constrict purely kinetic simulations due to small time step requirements for accurate collision modeling \cite{Turner_2013}. To address this constraint, hybrid approaches combine kinetic electron dynamics with a fluid dynamical treatment for ions and other heavy species \cite{Eremin_2016, Liu_2021, Klich_2022,Klich_2025, Vass_2024_2DHybr}. 

In this paper, simulations are performed using a hybrid simulation code that treats electrons kinetically while modeling heavy species, such as ions and neutrals, fluid dynamically. Simulations and experiments in COST-Jets operating in He/O$_2$ mixtures are combined and compared to explain electron dynamics and kinetic effects, particularly helium excitation through electron impact, and revisit the electron-heating-mode transition driven by variations in electronegativity by changing the O$_2$ admixture and applied voltage. Resolving electron kinetics can help interpret the experimental observations. The paper is organized as follows: Section 2 outlines the experimental setup and key parameters. Section 3 details the simulation, while Section 4 compares simulation results with phase-resolved optical emission spectroscopy (PROES) data from electronically excited helium discharges and analyzes the underlying physics in detail. Section 5 concludes the paper.

\section{COST-Jet -- Experimental setup}

\begin{figure}[H]
    \centering
    \makebox[0pt]{\includegraphics[scale=.45]{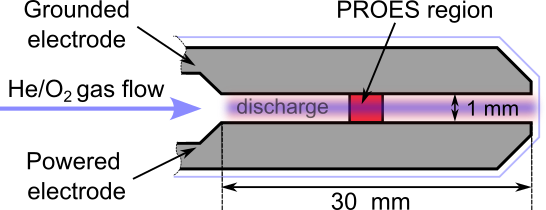}}
    \caption{Schematic representation of the COST-jet head. The domain highlighted in red indicates where PROES measurements were performed. }
    \label{COSTjet}
\end{figure}

The experiments utilize the COST reference microplasma jet, which is described in detail in previous works \cite{Korolov_2019, Korolov_2020HeMEta, Bischoff_2018TimeModulation,COSTRef_Golda_2016}. It consists of two parallel stainless-steel electrodes covered by two quartz plates, confining the discharge domain to \(1\,\text{mm} \times 1\,\text{mm} \times 30\,\text{mm}\), see Figure~\ref{COSTjet}. We use 5.0 and 4.5 purity grades of helium and oxygen gases, respectively. The flow rate of helium is kept constant at 1 slm, and the oxygen flow rate is varied between 0.5 and 5 sccm. To investigate the spatiotemporal dynamics of energetic electrons PROES measurements are conducted with spatial resolution perpendicular to the electrodes and nanosecond temporal resolution within the fundamental RF period.  

The experimental setup included a fast ICCD camera (4Picos) coupled with a 40 mm 4× CompactTL Telecentric Lens (Edmund Optics) and an interference filter centered at 700 nm (FWHM = 10 nm) to selectively observe the helium emission line at 706.5 nm. The ICCD covers the entire \(1\,\text{mm}\) electrode gap $L_{\rm{gap}}$ over $p_x=210\,$ pixels, generating a spatial resolution of approximately $4.8\,\mu\text{m/pixel}$. Temporal resolution was achieved by synchronizing the ICCD camera with the driving RF voltage waveform, enabling time-resolved measurements within the fundamental RF period using a camera gate width of 1 ns. The spatiotemporal emission intensities measured in the center position along the flow direction are then used to calculate the electron-impact excitation rate from the helium ground state to the He–I (1s3s) $^3$S$_1$ (22.7 eV) level. This calculation was performed using a collisional-radiative model with an effective radiative lifetime of around 6 ns. The geometry, gas handling system, voltage waveform control, and PROES diagnostics have been previously introduced and discussed in detail in \cite{Korolov_2019, Korolov_2020HeMEta, Bischoff_2018TimeModulation} and, therefore, will not be repeated here.

\section{Simulation method}

In this simulation model the discharge is simulated in 1D between the electrodes, while the neutral species are modeled in 2D to take their lateral transport into account. The hybrid approach addresses the widely disparate time scales in atmospheric pressure discharges -- from nanosecond electron dynamics to millisecond neutral evolution. To account for their non-equilibrium character \cite{Vass_2024_2DHybr, Eremin_2015Nonlocal}, electrons are treated kinetically, while ions and neutrals are modeled using fluid dynamical descriptions. This multi-domain strategy divides the computational problem into a plasma domain, where fast electron and ion processes are resolved, and a neutral domain, where convective and diffusive transport for neutrals are handled. Global convergence is achieved through a time-slicing algorithm \cite{Liu_2021, Kushner_2009TimeSlicing}. 

The kinetic treatments of electrons involves a PIC/MCC scheme \cite{Verboncoeur_2005, Birdsall, BastiTutorial, EduDonkó_2021, Wilczek_2015, Vass_2022PIC, Donkó_2009PIC, Wang_2021PIC}. To maintain numerical accuracy by resolving the Debye length  \(\lambda_D\), the Cartesian grid is chosen such that \(\lambda_D \ge 2\Delta x\), with the cell size defined by \(\Delta x = L_{\rm{gap}}/N_{\rm{x}}\) \cite{EduDonkó_2021,Turner_2013,Birdsall}. For this study, the electrode distance is \(L_{\rm{gap}} = 1\,\text{mm}\), which is resolved by \(N_{\rm{x}} = 201\) cells. Similarly, the time step is chosen to satisfy \(\omega_{pe}\Delta t \leq 0.2\) to resolve the electron plasma frequency. Additionally, high collisionality demands sufficiently small time steps to keep the collision probability per step low \cite{EduDonkó_2021,Turner_2013,Birdsall}. Therefore, \(N_{\rm{t}} = 10^6\) time steps are used to resolve the RF period of a driving frequency of \(f_{\rm{RF}} = 13.56\,\text{MHz}\). 

The background gas is helium with varying O$_2$ percentages. Electron impact cross sections for helium are taken from the LXCat (Biagi-v7.1) database \cite{LXCatproject, XSectionHelium}. In addition to elastic collisions and ionization, two electron‐impact excitation processes, from the ground state (He I) to the aggregated triplet state (19.82 eV) and singlet state (20.61 eV), are included, bringing the total number of electron‐impact processes with helium to four \cite{XSectionHelium}. This simulation adopts the previous approximation that $50\,\%$ of helium impact excitations result in the aggregate state of the He(2$^1$S) and He(2$^3$S) metastable states of helium, denoted He$^*$, capable of performing Penning ionization with O$_2$ \cite{Klich_2022,Klich_2025, Bischoff_2018TimeModulation, PICDonkó_2018}. The electron impact cross sections used in this work to describe collisions with oxygen are provided by the benchmark study of Gudmundsson \cite{Gudmundsson_2013}. A total of 15 distinct electron impact processes for O$_2$ are considered, covering elastic, dissociative, vibrational, rotational, electronic, and ionization processes. The following ion species are accounted for: He\(^{+}\), O\(_2^{+}\), O\(^{+}\), O\(_3^{-}\), O\(_2^{-}\), and O\(^{-}\). Small energy relaxation lengths at atmospheric pressure, combined with dominant friction, justify using the drift-diffusion approximation for ions \cite{Lieberman, Liu_2021, Klich_2022, Eremin_2016}. The electron mobility coefficients are functions of the reduced electric field. The diffusion coefficient is given by the Einstein relation \cite{Lieberman}. For helium ions, data is provided by Frost et al. \cite{Frost}, and for oxygen ions, data is obtained from Ellis et al. \cite{ELLIS1976177}. This formulation results in a convection-diffusion problem, which is solved using the Scharfetter–Gummel scheme \cite{ScharfetterGummel, VanDijkCFS}. 

At the electrodes, Dirichlet boundary conditions are applied to the densities of negatively charged species, i.$\,$e., \(n(x=0)=n(x=L_{\rm{gap}})=0\), while Neumann boundary conditions, i.$\,$e., \(\frac{\partial n}{\partial x}\Big|_{x=0}=\frac{\partial n}{\partial x}\Big|_{x=L_{\rm{gap}}}=0\) are used for positively charged species \cite{Vass_2024_2DHybr}. The ion-induced secondary electron emission coefficients for the positive ions are set as follows: \(\gamma_{\text{He}^{+}} = 0.2\), \(\gamma_{\text{O}_2^{+}} = 0.05\), and \(\gamma_{\text{O}^{+}} = 0.1\). 

Although helium metastables converge at the same timescales as electron impact processes, species like atomic oxygen need milliseconds to converge. In these cases, gas flow needs to be considered \cite{Liu_2021}. Therefore, a two-dimensional convection-diffusion-reaction equation governs the transport of neutral species to account for the mass flow toward the jet outlet. Transport between the quartz plates of the jet is neglected. Fick's law considers the neutral transport perpendicular to the electrodes \cite{Waskoenig_2010AtomicOxigen}. Surface loss probabilities, which depend on the neutral species, are applied at the electrodes \cite{Liu_2021}. In the gas flow direction along the jet head, diffusive transport is negligible, and only a convective flow field is considered \cite{Liu_2021}. Due to the low inflow rate and the resulting low Mach number \cite{Liu_2023}, the gas flow remains laminar with negligible compressibility. Since the plasma does not significantly disturb the flow, the velocity field can be computed once using a Hagen-Poiseuille (parabolic) profile \cite{bartelmann2014theoretische}. The governing equation is solved numerically using a Crank-Nicolson scheme. Details on the specific reactions are provided in \cite{Liu_2021, TurnerPaper_2016}. The simulation in this study accounts for 92 chemical reactions involving electrons, ions, and neutral species. The considered neutral species include He\(^{*}\), O, O(\(^{1}\)D), O\(_2\) (\(v = 1\)–\(4\)), O\(_2\)(\(a^1\Delta_g\)), O\(_2\)(\(b^1\Sigma_g\)), O\(_3\), and O\(_3\)(\(v\)). Neutral species exhibit a broad range of convergence times due to their different production and loss processes. 

\section{Results} 

The following section analyzes the excitation dynamics under varying O$_2$ admixtures and voltages, both experimentally and numerically.

\begin{figure*}[htbp]
    \centering
    \makebox[0pt]{\includegraphics[width=1.0\textwidth]{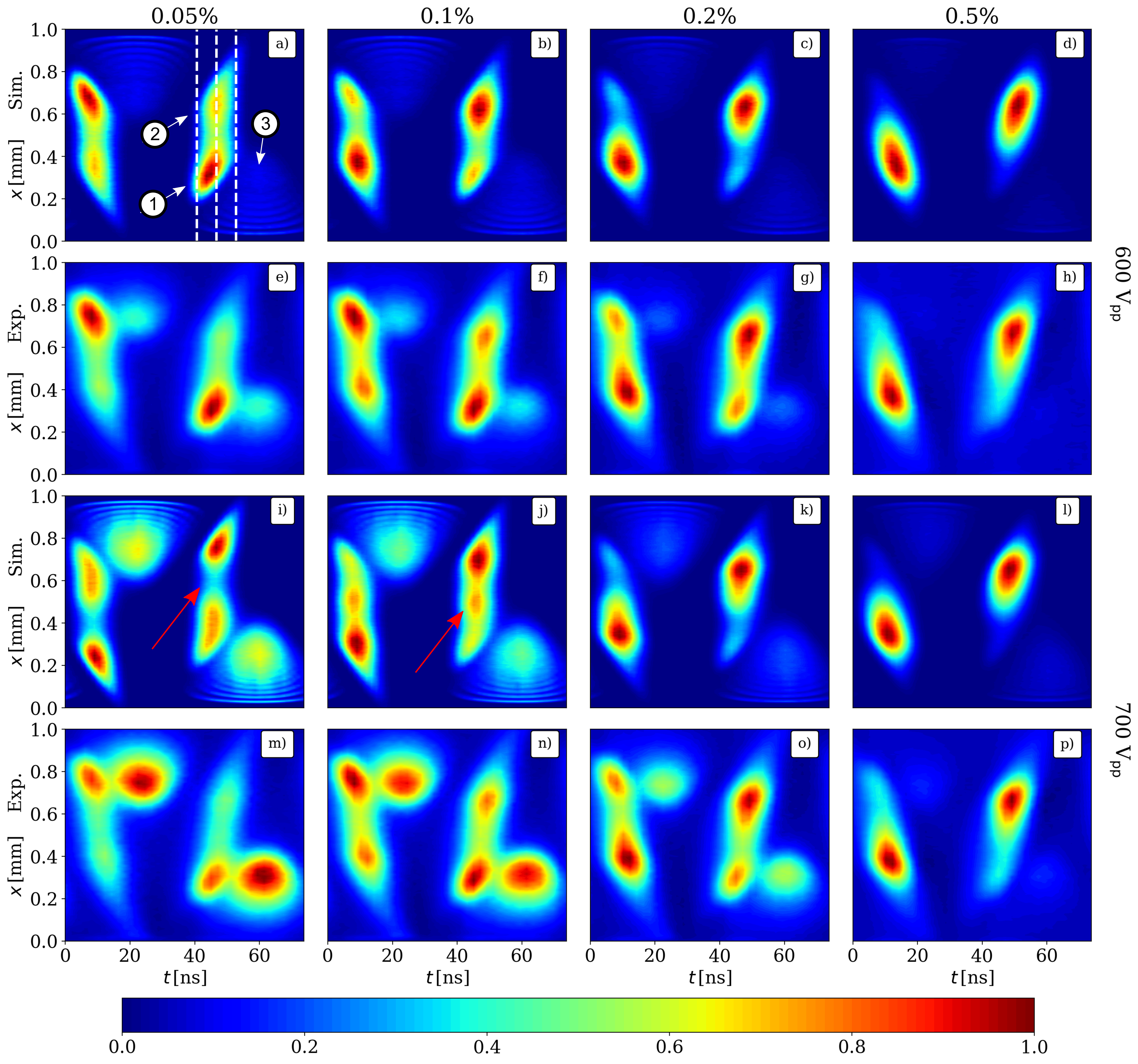}}
    \caption{Production rate of He\(^{*}\) (simulations) and electron impact excitation rate of He–I (1s3s) \(^{3}\)S\(_1\) (experiments) for O\(_2\) admixtures of $0.05\,\%$, $0.1\,\%$, $0.2\,\%$, and $0.5\,\%$. Top two rows correspond to \(V_{\rm{pp}} = 600\,\text{V}\) and bottom two to \(V_{\rm{pp}} = 700\,\text{V}\) (odd rows: simulations; even rows: experiments). White vertical lines at \(39.8\,\text{ns}\), \(45.7\,\text{ns}\), and \(51.6\,\text{ns}\) mark peak ohmic dynamics, while markings ‘1’, ‘2’, and ‘3’ indicate regions of increased excitation. Discharge conditions: \( f_{\rm{RF}} = 13.56\,\text{MHz} \), \( p = 10^5\,\text{Pa} \), and \( L_{\rm{gap}} = 1\,\text{mm} \).}
    \label{ExecVariation}
\end{figure*}

\myref{ExecVariation} shows the spatiotemporal excitation rate of the helium ground state to the He–I (1s3s) $^3$S$_1$ level ($22.7\,\rm{eV}$) observed in the experiments. The experimental data is compared with the results obtained from the hybrid simulation. Simulations use He$^*$ production rates to represent the 1s3s excitation, providing a qualitative match despite their lower energy thresholds. The two upper rows correspond to simulations and experiments conducted with a peak-to-peak voltage of \( V_{\rm{pp}} = 600\,\text{V} \), while the bottom two rows correspond to a peak-to-peak voltage of \( V_{\rm{pp}} = 700\,\text{V} \). The simulation results are displayed on rows one and three, while the experimental results are shown on rows two and four. The columns are organized by specific O$_2$ admixtures, with values of $0.05\,\%$, $0.1\,\%$, $0.2\,\%$, and $0.5\,\%$ from left to right. By varying voltage and O$_2$ admixture, the transition from Penning‐Gamma to ohmic heating modes is induced. Ohmic heating stems from a strong bulk field driven by low DC conductivity \cite{Lieberman},
\begin{align}
\sigma_{\text{DC}} = \frac{e^2 n_{\rm{e}}}{m_{\rm{e}} \nu_{\rm{m}}},
\label{dc_conductivity}
\end{align}
where \(e\) is the elementary charge, $n_{\rm{e}}$ the electron density, \(m_{\rm{e}}\) is the electron mass, and \(\nu_{\rm{m}}\) is the frequency of momentum transfer of electrons. At atmospheric pressure, frequent electron–neutral collisions lower $\sigma_{\rm DC}$, which requires a strong bulk electric field to maintain current continuity \cite{SchulzeDriftAmbiMode}. Adding O$_2$ creates negative ions and $n_{\rm e}$ is reduced due to additional negative charges. Subsequently, $\sigma_{\rm DC}$ also reduces, and the electric field is enhanced \cite{Hemke_2013OOhmicMode, Liu_2021ModeTransition}. In the Penning-Gamma mode, electrons are accelerated by the strong sheath electric field. These electrons mainly collide with the helium background due to its higher gas fraction compared to O$_2$. Helium metastables are formed, which trigger Penning ionization with O$_2$, generating locally an electron-ion pair. The electron can create additional metastables while the ion gets accelerated to the electrode, which can cause the emission of secondary electrons from the surface \cite{Bischoff_2018TimeModulation,Liu_2021ModeTransition,RoleOFHEliumMEtaCostJet}.

\myref{ExecVariation} panel a) establishes three regions with increased electron-impact excitation of helium during the second half-period of the driving voltage signal: (1) peak '1' near the driven electrode, (2) peak '2' in the vicinity of the grounded electrode, and (3) region '3' at the electrode, where Franck–Hertz-like striations indicate a Penning–Gamma contribution, while '1' and '2' indicate ohmic contribution \cite{Bischoff_2018TimeModulation, FranckHertz}. At \(V_{\rm{pp}} = 600\,\text{V} \), experiments and simulations show good agreement. For $0.05\,\%$ O$_2$ admixture, peak '1' exhibits the highest intensity in both cases, while peak '2' is clearly visible in simulations but only faintly detectable in experiments. Penning-Gamma contribution (peak '3') progressively vanishes with increasing O$_2$ admixture in both cases. At 0.1$\,\%$  O$_2$ admixture, ohmic contribution is dominant for experiments and simulations. Simulations (panel b)) show enhanced relative intensity of peak '2' compared to peak '1'. At the same time, experiments show this transition from a dominant peak '1' to a dominant peak '2' occurring at higher O$_2$ concentrations (a delayed shift). Both simulations and experiments converge to a singular maximum at peak '2' by 0.5$\,\%$ O$_2$ admixture.\par 
\myref{ExecVariation} panels i)–l) (simulations) and m)–p) (experiments) present admixture variation results at \( V_{\rm{pp}} = 700\,\text{V} \). Compared to \( V_{\rm{pp}} = 600\,\text{V} \), the Penning-Gamma mode contribution intensifies in both cases due to enhanced sheath electric fields, then diminishes with increasing O$_2$ admixture \cite{Liu_2021ModeTransition}. At 0.2$\,\%$  and 0.5$\,\%$  O$_2$ admixture (panels k) – l) simulation, o) – p) experiment), both datasets exhibit similar electron impact excitation dynamics (peaks '1' and '2') in comparison to  \( V_{\rm{pp}} = 600\,\text{V} \). For $0.05\,\%$ and 0.1$\,\%$ O$_2$ admixture, the simulations (panels i and j) differ from the experimental data (panels m and n). An additional dark space in panel i) and an additional excitation peak in panel j) are visible, both marked by red arrows. The shift from peak '1' to peak '2', visible for both voltages, was previously observed only experimentally and not in the complementary simulations \cite{Liu_2021ModeTransition}. Due to the kinetic treatment of electrons in the presented simulation, these features are also present in the results obtained by the simulation used in this paper.

Several factors may contribute to the slight discrepancies in the Penning-Gamma mode peak and the admixture-dependent shift in mode transition between the experiment and the simulation. First, the limited optical resolution could explain the loss of striation details in the experimental data. Second, the voltage probe’s inherent accuracy of ~5\,\% may introduce deviations of approximately 30$\,\rm{V}$, altering discharge conditions, influencing secondary electron emission, and the electric field. Finally, uncertainties in gap distance, up to $50\,\mu \rm{m}$, could further affect the results. The differences in intensity in the simulation can be attributed to the choice of secondary electron emission coefficients, which significantly influence the Penning-Gamma mode. Determining these coefficients is challenging and could account for deviations from experimental observations \cite{SEEYIELD,Noesges_2023SEE,Sun_2020SEE,Derzsi_2015SEE,Daksha_2017SEE,Kawamura_2014}. 

Furthermore, the one-dimensional plasma model cannot capture inherently three-dimensional effects, such as flow variations near the quartz walls or the changing plasma profile along the discharge channel. Electronegativity falls continuously along the jet as negative ions are depleted by recombination and detachment. This one-dimensional simulation, which omits axial plasma transport and field gradients cannot reproduce this continuous change in electronegativity \cite{Vass_2024_2DHybr}. Performing a simulation at an axial point where the experiment and the simulation display the same degree of electronegativity might align the transition of excitation maxima closer, depending on the O$_2$ admixture.

All these effects can contribute to the discrepancies between the simulation and the experiment. Despite these slight localized discrepancies, the overall trends in simulation and experiment agree closely, showing that the one-dimensional hybrid model reliably captures the electron‐kinetic effects during a mode transition and the intensity shift from peak '1' to peak '2'. Although tuning parameters (e.g.\ secondary‐electron yield or gap distance) could yield different results, it is not the scope of this work. The current level of accuracy is sufficient to use the simulation results for a detailed analysis of the excitation dynamics of the discharge. To better understand the features of the impact excitation during the transition, the electric field is examined.

\begin{figure*}[htbp]
    \centering
     \makebox[0pt]{\includegraphics[width=1.0\textwidth]{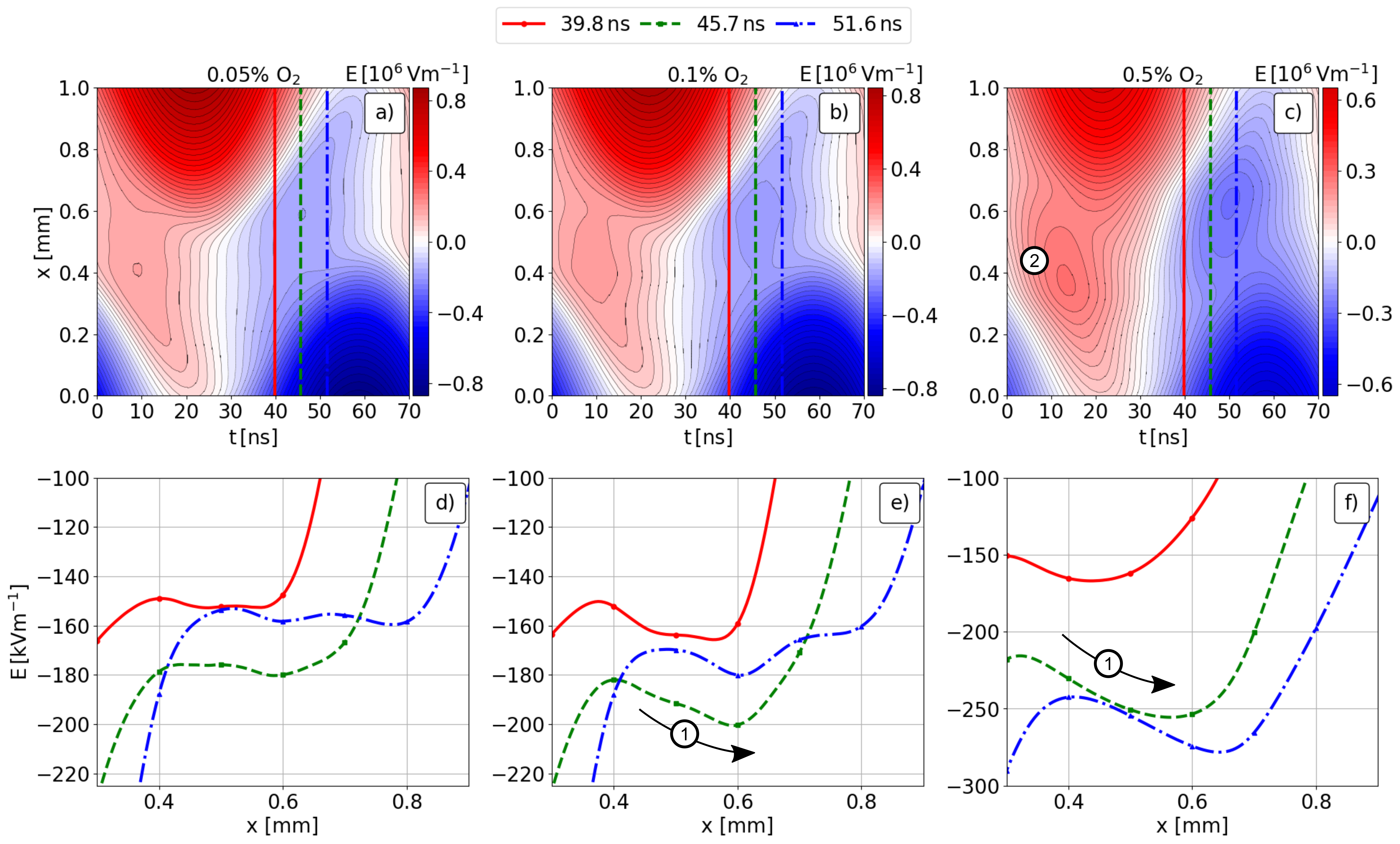}}
    \caption{Electric field $E$ for O\(_2\) admixtures of a) $0.05\,\%$, b) 0.1\%, and (c) 0.5\%. Snapshots of $E$ for \(t_1 = 39.8\,\text{ns}\), \(t_2 = 45.7\,\text{ns}\), and \(t_3 = 51.6\,\text{ns}\) in d) - f) corresponding to vertical lines in a) - c). Markings: '1' - electric field gradient; '2' - electric field 'hump'. Discharge conditions: \(f_{\rm{RF}} = 13.56\,\text{MHz}\), \(V_{\rm{pp}} = 600\,\text{V}\), \(p = 10^5\,\text{Pa}\), \(L_{\rm{gap}} = 1\,\text{mm} \).}
    \label{field_snaps1}
\end{figure*}

The ohmic electric field in the central region dominates metastable production and must be examined, as a locally inhomogeneous electric field may cause an inhomogeneous excitation profile. Therefore, \myref{field_snaps1} shows the electric field for a driving voltage of \( V_{\text{pp}} = 600\,\text{V} \) for varying O$_2$ admixtures. The top row displays the spatiotemporal distribution of the electric field. Vertical lines in panels a)–c) indicate snapshots at $39.8 \, \text{ns}$ (red), $45.7 \, \text{ns}$ (green), and $51.6 \, \text{ns}$ (blue), chosen for their pronounced electric-field dynamics. The bottom rows (d) to (f)) show the corresponding snapshots of the electric field. The focus is on the area from \( x = 0.3\,\text{mm} \) to \( x = 0.9\,\text{mm} \). From a) to c), with increasing O$_2$ admixture, the spatiotemporal dynamics increase in the center of the electric field, comparable to the dynamics seen in the electron impact excitation rate. A local electric field 'hump' in form of a field gradient reversal emerges when the O$_2$ admixture is increased from $0.05\,\%$ to \(0.1\%\) (at \(x = 0.6\,\mathrm{mm}\) between \(45.7 \, \text{ns}\) and \(51.6 \, \text{ns}\)). Increasing the O$_2$ admixture to $0.5\%$, the gradient of the 'hump' increases in magnitude as does the electric field magnitude, as shown in \myref{field_snaps1} f) (black arrows marked '1' indicate moments with steep spatial change). Looking at \myref{field_snaps1} panel c), one distinct 'hump' is observed at the point marked by '2'. Following the field gradient change, the excitation rate of helium by electron impact exhibits also a single dominant 'hump', consistent with experimental observations and the simulations. Typically, although the peak electric field decreases in magnitude, the bulk electric field increases with rising electronegativity \cite{SchulzeDriftAmbiMode}.

\begin{figure*}[htbp]
    \centering
     \makebox[0pt]{\includegraphics[width=1.0\textwidth]{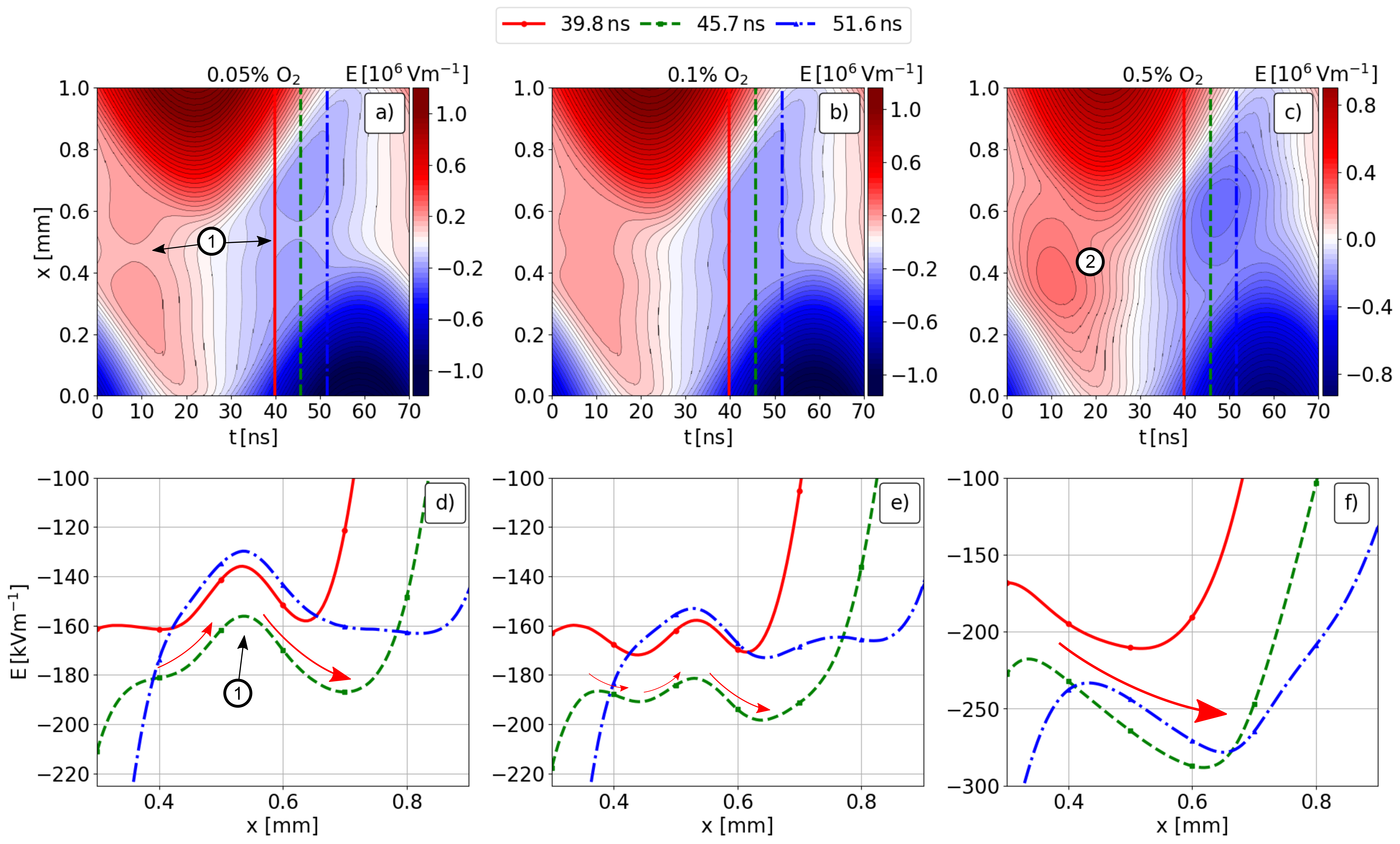}}
    \caption{Electric field $E$ for O\(_2\) admixtures of a) $0.05\,\%$, b) 0.1\%, and (c) 0.5\%. Snapshots of $E$ for \(t_1 = 39.8\,\text{ns}\), \(t_2 = 45.7\,\text{ns}\), and \(t_3 = 51.6\,\text{ns}\) in d) - f) corresponding to vertical lines in a) - c). Markings: '1' and '2' - electric field 'humps'; red arrows - electric field gradients. Discharge conditions: \(f_{\rm{RF}} = 13.56\,\text{MHz}\), \(V_{\rm{pp}} = 700\,\text{V}\), \(p = 10^5\,\text{Pa}\), \(L_{\rm{gap}} = 1\,\text{mm} \).}
    \label{field_snaps2}
\end{figure*}

In order to better understand also the dynamics of excitation at \( V_{\rm{pp}} = 700\,\text{V} \), the corresponding electric field is shown in  \myref{field_snaps2}. The figure structure has the same layout as \myref{field_snaps1}. In panels a) and d) for an O$_2$ admixture of $0.05\,\%$, modulation at point '1' in the bulk electric field is visible. At this point, a field 'hump' in the opposite direction to the previous emerges. Locally lower field means less momentum gain for electrons explaining the dark space in \myref{ExecVariation} i). The red arrows highlight the spatial electric field changes. For an O$_2$ admixture of \(0.1\%\) (panels b), e)), additional local maxima emerge, accompanied by three consecutive electric field gradients (red arrows). Compared with the excitation rates in \myref{ExecVariation} j), 'hump' in the electron impact excitation of helium can also be observed at these positions. Each time a local increase in an electric field is observed, a peak in the excitation of He$^*$ can be detected. At \(0.5\%\) O$_2$ admixture, again, only a single well-defined electric field 'hump'  marked by '2' is visible in \myref{field_snaps2} c). Similarly to this field structure, only one peak in the excitation rate is visible in \myref{ExecVariation} l).

The electric field humps are spatiotemporally aligned with the peaks in the helium excitation rate. Due to the locality of atmospheric pressure plasmas, electron-impact processes are primarily driven by the ohmic bulk electric field \cite{Vass_2021HEN2_Jet,Eremin_2015Nonlocal}. With increasing electronegativity, both the electric field humps and their associated gradients become more pronounced, along with an increase in the bulk electric field. Local reversals in the electric field gradient reveal that during the transport of charged species in an RF cycle, a switch from positive to negative charge density occurs at points where the gradient changes. The plasma properties are studied in greater detail in the following to understand the occurrence of these gradient reversals.


In the following, plasma properties are examined in more detail to understand the behavior of charged species and the effect of O$_2$ admixture on composition and electronegativity.
\begin{figure*}[htbp]
    \centering
     \makebox[0pt]{\includegraphics[width=1.0\textwidth]{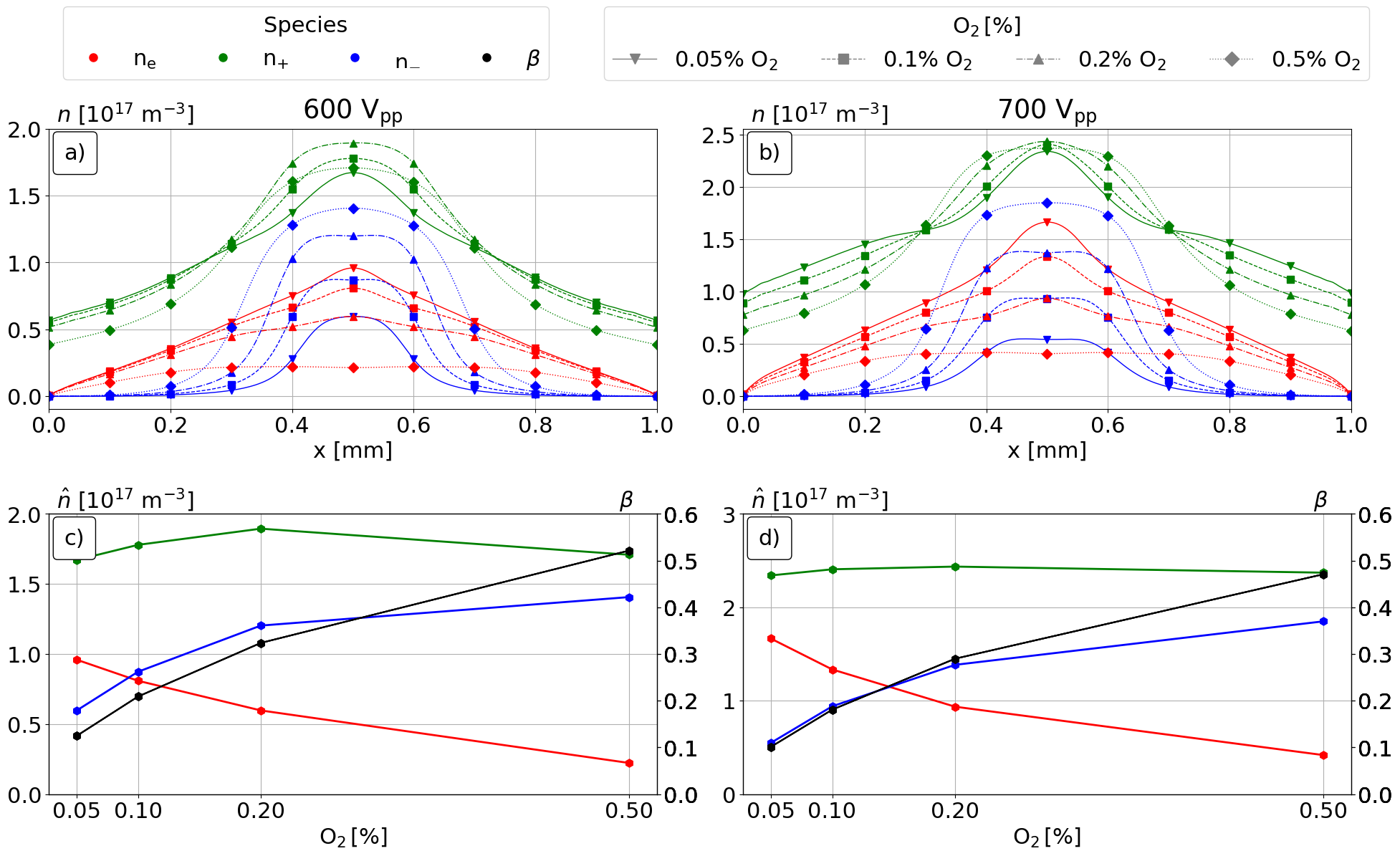}}
\caption{Top row: Time-averaged densities — electrons $n_{\rm{e}}$ (red), positive ions $n_{\rm{+}}$ (green), negative ions $n_{\rm{-}}$ (blue). Bottom row: Peak densities of $n_{\rm{e}}$, $n_{\rm{+}}$, $n_{\rm{-}}$ and electronegativity $\beta$ (black). Conditions: Left: \(V_{\rm{pp}} = 600\,\text{V}\); Right: \(V_{\rm{pp}} = 700\,\text{V}\); \(f_{\rm{RF}} = 13.56\,\text{MHz}\), \(p = 10^5\,\text{Pa}\), \(L_{\rm{gap}} = 1\,\text{mm}\).
}\label{dens}
\end{figure*}
\myref{dens} compares simulation results at \(V_{\rm{pp}} = 600\,\text{V}\) (left) and \(V_{\rm{pp}} = 700\,\text{V}\) (right). The top row (a), b)) shows electron profiles (\(n_{\rm{e}}\), red), positive ions (\(n_+\), green) and negative ions (\(n_-\) throughout the electrode gap, while the bottom row (c),d)) shows their peak values \(\hat{n}_+\), \(\hat{n}_{\rm e}\) and \(\hat{n}_-\), and the electronegativity \( \beta = \overline{n_{-} }/\, \overline{n_{\rm{e}}} \)  versus O$_2$ admixture, with densities on the left axis and $\beta$ on the right. Increasing the O\(_2\) admixture leads to a pronounced central peak in \(n_-\), while \(n_{\rm e}\) decreases and broadens, \(\hat{n}_+\) remains nearly constant \cite{Liu_2021ModeTransition, DynElecNegORF, Derzsi_2022Electronegative}. These trends persist in both \(V_{\rm{pp}} = 600\,\text{V}\) and \(700\,\text{V}\). The peak densities of \(\hat{n}_{\rm e}\) decrease with increasing admixture, while that of \(\hat{n}_+\) increases, as does $\beta$. The peak value of \(\hat{n}_+\) remains largely the same. As a result, the overall charge density diminishes and the plasma approaches quasineutrality in this regime. O$_2$ admixture and, therefore, the degree of electronegativity influences space charge dynamics, which may cause local electric-field variations. Examining these charge density changes is fundamental for understanding discharge behavior. Therefore, these dynamics will be inspected in more detail.


\begin{figure*}[htbp]
    \centering
    \includegraphics[width=1.0\textwidth]{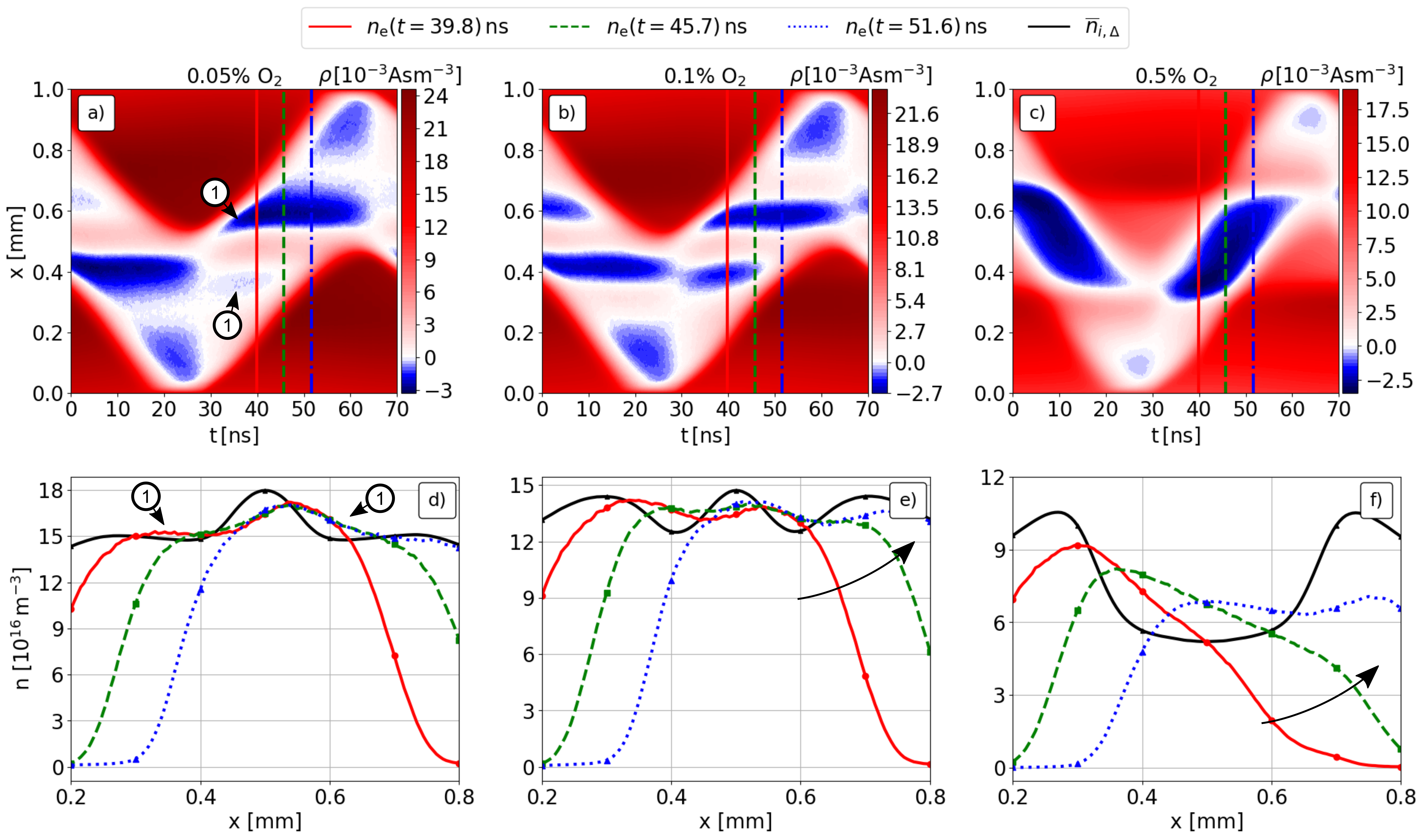}
    \caption{Charge density $\rho$ for O\(_2\) admixtures of a) $0.05\,\%$, b) 0.1\%, and (c) 0.5\%. Snapshots of $n_{\rm{e}}$ for \(t_1 = 39.8\,\text{ns}\), \(t_2 = 45.7\,\text{ns}\), and \(t_3 = 51.6\,\text{ns}\) and net positive ion density $\overline{n}_{i,\Delta}$ in d) - f) corresponding to vertical lines in a) - c). Discharge conditions: $f_{\rm{RF}} = 13.56\,\text{MHz}$, $V_{\rm{pp}} = 700\,\text{V}$, $p = 10^5\,\text{Pa}$, $L_{\rm{gap}} = 1\,\text{mm}$.}
    \label{chargedens}
\end{figure*}

The mode transition at $V_{\rm{pp}} = 600\,\text{V}$ and $V_{\rm{pp}} = 700\,\text{V}$ display fundamentally the same physical processes. The study performed at $700\,\text{V}$ displays additional features compared to the experiments. In order to understand these features and the field behaviour during the mode transition/  excitation-maxima shift, the focus is on the study conducted at  $700\,\text{V}$. A local changing charge density must govern the resulting local changes  in the electric field in the form:
\begin{align}
    \frac{\partial E}{\partial x} \propto \rho.
    \label{dens_grad}
\end{align}
Therefore, \myref{chargedens} displays, in the top row, the spatiotemporal charge density $\rho$. The oxygen concentration increases column-wise, analogous to \myref{field_snaps1}. Snapshots at $39.8\,\text{ns}$ (red, solid), $45.7\,\text{ns}$ (green, dashed), and $51.6\,\text{ns}$ (blue, dashed-dotted), marked in panels a) to c) by vertical lines, correspond to times when the excitation dynamics are most pronounced. The associated snapshot profiles for the electron density $n_{\rm{e}}$ are shown in the bottom row (panels d) to f)) in the corresponding colors and styles. The time-averaged net positive ion density is introduced and defined as $\overline{n}_{i,\Delta} = \left< n_{+} - n_{-} \right>_t$ and displayed in black in the bottom row panels. The low mobility of all heavy charged species compared to electrons justifies time averaging. The focus is on the region from $0.3\,\text{mm}$ to $0.9\,\text{mm}$. 

In the bottom row, net positive ion and electron dynamics resemble at first glance the non-neutral regime observed in electropositive gases \cite{Klich_2022}. These electropositive discharges feature a quasi-static ion distribution far exceeding the electron density distribution in the center (creating the non-neutral region) and a dynamic electron soliton density profile traversing the discharge gap \cite{Klich_2022, Vass_2021HEN2_Jet, Elbadawy_2024Soliton}. However, in the electronegative He/O$_2$ mixture, the discharge center is much closer to quasineutrality because negative ions impact the overall net positive charges $\overline{n}_{i,\Delta}$ in the center, effectively reducing it. When the electron profile, here only loosely describable as a solitary structure, traverses the discharge gap, the electron profile may exceed $\overline{n}_{i,\Delta}$ and create negative charge density patches and local changes in the electric field (see eq. \ref{dens_grad}). These instances are marked with "1" in \myref{chargedens}. The electron profile experiences an inhomogeneous field in the center and distorts during traversal.

At $t_1$ (see \myref{chargedens} panels a) and d)), for $0.05\,\%$ O$_2$ the electron density $n_{\rm{e}}$ slightly exceeds the averaged ion density $\overline{n}_{i,\Delta}$ at $0.35\,\mathrm{mm}$, creating a negative space-charge patch. Immediately afterwards, at $0.5\,\mathrm{mm}$, where \(\overline{n}_{i,\Delta}\) reaches its peak, a positive patch forms. This positive patch enhances the local electric field and electron acceleration. Between $0.6\,\mathrm{mm}$ and $0.7\,\mathrm{mm}$, the situation changes again when a negative space‐charge zone forms, reducing the field (see inverted hump in \myref{field_snaps2} at $0.05\%$ O$_2$) and limiting electron momentum gain. For an O$_2$ admixture of $0.1\,\%$ (panel b)), the negative ion density confined to the discharge center increases, resulting in a reduced peak of $\overline{n}_{i,\Delta}$ at $0.5\,\rm{mm}$ (see panel e)). Similar charge density patches as before persist, but the first patch at $0.35\,\rm{mm}$ intensifies. Here, $n_{\rm{e}}$ exceeds $\overline{n}_{i,\Delta}$ more pronounced. Due to the reduced peak of $\overline{n}_{i,\Delta}$ at $0.5\,\rm{mm}$, the subsequent positive charge density patch becomes less noticeable. These local variations in charge density lead to spatial inhomogeneities in the bulk electric field (see eq. \ref{dens_grad} and \myref{field_snaps2} at $0.1\,\%$ O$_2$). Electrons experience varying momentum gain from the inhomogeneous electric field during traversal. From $t_1$ to $t_3$, the electron density profile undergoes a sloshing motion (indicated by the black arrow in \myref{chargedens} panel e)) while traversing the discharge gap. Further increasing the O$_2$ admixture to $0.1\,\%$ creates a depletion region of net positive charge in $\overline{n}_{i,\Delta}$, extending at approximately $0.3\,\rm{mm}-0.7\,\rm{mm}$ (panel f)). In this region, the electron density exceeds $\overline{n}_{i,\Delta}$ at all times from $t_1-t_3$, producing a negative space charge zone (panel c)). Consequently, a localized enhancement of the electric field occurs, which increases electron momentum gain and amplifies the electron sloshing motion (indicated by the black arrow in panel f)).

After the inspection of electron impact excitation, electric field, and the dynamics of charges, a clearer picture emerges for the physical mechanism: Increasing the electronegativity with an increase of O$_2$ admixture produces more negative ions. These negative ions lead to a reduction of the positive net charge of ions in the center. In turn, electrons can create localized negative space charge zones during the traversal of the electron density profile of the discharge gap. In addition to the inherently high bulk electric field of atmospheric pressure plasma jets, the electric field can be enhanced or reduced due to charge density patches in the bulk area. Locally generated electrons traversing this region of enhanced electric field acquire additional momentum proportional to the degree of field enhancement. This local enhancement determines the relative intensity of the production rate of He$^*$.


\section{Summary and Conclusion}

This work studies the electron heating mode transition experimentally and numerically in He/O$_2$ mixtures in the COST-Jet. The investigation focused on electron dynamics and their sensitivity to electronegativity, which in turn governs the excitation of helium by electron impact.

This work employs a time-sliced hybrid code to model discharges in He/O$_2$ mixtures in the COST-Jet. The simulation treats electrons kinetically via PIC/MCC, while heavy species (ions and neutrals) are modeled fluid dynamically. The governing equations for ions and neutrals are continuity equations in drift diffusion approximations. Charged species are resolved in one dimension, and neutral transport is captured in two dimensions to account for the gas flow. This approach efficiently bridges the disparate time scales of electron, ion, and neutral dynamics, enabling a detailed analysis of electron kinetics and the electron heating mode transition induced by electronegativity. 

The simulation results provide insights into the electron kinetic effects during electronegativity-induced heating mode transition, specifically during the phase where the ohmic mode contribution was dominant. Penning–Gamma and ohmic electron‐heating contributions appear in experiments and simulations. This work focused on the features in the electron impact excitation rate during bulk traversal of electrons driven primarily by the ohmic field. The ohmic mode, characterized by a high bulk electric field from a high degree of collisionality and electronegativity, is the main contributor to the discharges shown in this paper operated in the given conditions. In contrast to electropositive discharges \cite{Klich_2022}, in the He/O$_2$ cases, the ohmic mode heating contribution displays an additional intensity peak at the collapsing boundary sheath \cite{Bischoff_2018TimeModulation, Klich_2022}. Spatial variations of the electric field, i.$\,$e. reversals in the electric field gradient and locally enhanced electric fields in the form of 'humps', in the discharge center were found to influence electron dynamics. These spatial inhomogeneities in the electric field are found to be responsible for creating the additional electron impact excitation peak. Negative ions reduce the total positive space charge in the bulk region, allowing electrons to exceed the local positive space charge density and create negative space charge zones. These localized negative space charge zones cause the local enhancement in the electric field, reversing the electric field gradient and affecting the electron momentum gain. 

The influence of electronegativity on the bulk electric field and the spatial distribution of net positive ion background and electrons was further demonstrated by varying the O$_2$ admixture concentration. Increased O$_2$ leads to higher electronegativity, i.$\,$e. more negative ions and less electrons. This results in an overall reduction of charge density in the discharge. Consequently, more pronounced negative space charge zones form during the electron density profile traversal phase over the discharge gap, resulting in significant inhomogeneities in the center electric field and more substantial gradient changes. The increase of O$_2$ admixture shifts the dominant peak of the helium electron impact excitation from the expanding to the collapsing sheath by producing local negative space‐charge zones (where $n_e>\overline{n}_{i,\Delta}$) that reverse the field gradient and enhance the electric field locally, accelerating electrons toward the collapsing sheath. The simulation results closely match experimental measurements, therefore providing a reliable explanation for the features observed in the PROES measurements.

\section*{Acknowledgments}
This work was supported by the Deutsche Forschungsgemeinschaft (DFG) within the framework of the Collaborative Research Center SFB 1316, Projects A4, A5 and A8 as well as Research Grant MU 2332/12-1. 
\printbibliography
\end{document}